\documentclass[14pt,a4paper]{article}
\usepackage{amssymb}
\usepackage{amsfonts,amsmath,amsxtra}
\usepackage{graphicx}
\usepackage{lscape,epsf}
\usepackage[bookmarks=true]{hyperref}

\allowdisplaybreaks[1]

\newcommand{\eps}{\epsilon}

\renewcommand{\t}{\tau}

\newcommand{\Ic}{\mathcal{I}}
\newcommand{\Nc}{\mathcal{N}}

\newcommand{\form}[1]{\mathbf{{#1}}}

\newcommand{\I}{\form{I}}
\newcommand{\Z}{\form{Z}}
\renewcommand{\S}{\form{S}}
\newcommand{\T}{\form{T}}
\newcommand{\U}{\form{U}}

\newcommand{\W}{\form{W}}
\newcommand{\V}{\form{V}}

\newcommand{\Wt}{\form{\tilde{W}}}
\newcommand{\Vt}{\form{\tilde{V}}}

\newcommand{\h}[1]{h^{#1}}

\newcommand{\sg}{\sqrt{-G}}
\newcommand{\wg}{\wedge}
\newcommand{\lpl}{\square_4}

\newcommand{\mt}{\tilde{m}}

\newcommand{\tr}{\mbox{\,Tr\,}}

\newcommand{\pr}{\partial}
\renewcommand{\d}{\mathrm{d}}

\def\uor{U(1)_{\cal{R}}}

\def\ds{\displaystyle}
\def\bea{\begin{array}{c}}
\def\ea{\end{array}}
\def\be{\begin{equation}\bea\ds}
\def\ee{\ea\end{equation}}

\def\KS{Klebanov-Strassler }

\begin{document}

\title{On $\Ic$-even Singlet Glueballs in the Klebanov-Strassler Theory}
\author{  Ivan Gordeli${}^{\,a}$ and Dmitry Melnikov${}^{\,b,\,c}$}
\date{}
\maketitle
\thispagestyle{empty}
\vspace{0.5cm}
\begin{center}
\itshape
${}^{a}$ University of Minnesota, School of Physics and Astronomy,\\
116 Church St. SE, Minneapolis, MN 55455, USA
\\[1.mm]
$^{b}$ Raymond and Beverly Sackler School of Physics and Astronomy,\\
Tel Aviv University, Ramat Aviv 69978, Israel
\\[1.mm]
${}^{c}$ Institute for Theoretical and Experimental Physics,\\
B.~Cheremushkinskaya, Moscow 117259, Russia
\\[1.mm]
\end{center}

\vspace{-11cm}
\begin{flushright}
{TAUP-2908/09}\\
{ITEP-TH-94/09}

\end{flushright}
\vspace{10.5cm}

\begin{abstract}
In this note we study vector fluctuations over the  Klebanov-Strassler type IIB supergravity solution that are even under the $\Ic$ conjugation. We are interested only in the states invariant under the global  $SU(2)\times SU(2)$ symmetry. Apart from the glueball dual to the $\uor$ current there is one more $1^{++}$ state, which is a member of a massive vector multiplet containing also a scalar $0^{++}$. Combined with previous results, our analysis allows to complete the list of low energy singlet supermultiplets in the Klebanov-Strassler theory.
\end{abstract}

\newpage


\section{Introduction}

The Klebanov-Strassler (KS) theory~\cite{KS} is one of the most interesting examples of the holographic correspondence~\cite{AdS/CFT}. It provides the dual description of a $\Nc=1$ supersymmetric non-conformal gauge theory with dynamical chiral symmetry breaking and confinement. A relatively simple construction from the point of view of the correspondence, it is often proposed to model the real world physics in such areas as viable string compactifications, inflation, supersymmetry breaking and more (e.g. in~\cite{KS-examples}).

Our motivation for working with the KS solution is the analysis of the low energy degrees of freedom (glueballs) of a theory that is a close relative of the pure gauge $\Nc=1$ supersymmetric Yang-Mills theory. Holographic methods applied to such a class of theories could provide qualitative estimates for physical observables in the strong coupling regime. In particular such estimates would be useful in the setup discussed in~\cite{JMS}, where we are interested in computing transition matrix elements of operators in the strongly coupled hidden sector~\cite{hv}. It is also interesting to test the holography against the theoretical predictions on glueballs in QCD~\cite{Charmonium}. For further motivation for calculation of glueball spectra using holographic correspondence see~\cite{Glueballs-motivation}.

In this work we continue a series of earlier publications on the spectrum of glueballs in the KS theory~\cite{GHK}--\cite{DMS}. We are interested in the singlets with respect to the global $SU(2)\times SU(2)$ symmetry, since this sector contains the states of the pure gauge $\Nc=1$ theory. We only study bosonic states, since the spectrum of fermions is automatically restored by supersymmetry.

As usual the glueballs are classified by the $J^{PC}$ quantum numbers. The value of $C$ can be determined as follows. As one learns from~\cite{GHK,KW} there is a $\mathbb{Z}_2$ symmetry of the KS solution, which is realized as the swap of the two $S^2$ in the base of the conifold followed by flipping the signs of the $F_3$ and $H_3$ forms. In the dual gauge theory it corresponds to the interchange of the bifundamental fields $A_i\leftrightarrow B_i$ combined with the charge conjugation. Therefore this symmetry, called the $\Ic$-symmetry, determines the $C$ conjugation properties of the bulk modes.

The glueball sector odd under the $\Ic$ conjugation was completely studied in~\cite{BDKS,DMS}. In the $\Ic$-even sector the spectrum of $0^{++}$ states was computed in~\cite{BHM1,BHM2} and the graviton multiplet containing $2^{++}$ and $1^{++}$ glueballs was studied in~\cite{DM}. In this paper we find another $1^{++}$ glueball in this sector, which turns out to be degenerate with one of the seven $0^{++}$, and show that there are no other singlet vector states. Together with the previous results this allows us to conclude that all singlet supermultiplets are known and we propose the full spectrum of singlet glueballs in the KS theory. The only missing states are the six pseudoscalars $0^{-+}$ that are members of the scalar multiplets together with the six $0^{++}$. The pseudoscalar states will be discussed in the future work~\cite{finale}.

A comment on the spectrum of $0^{++}$ states is required. The seven equations for those states form a complicated coupled system~\cite{BHM1}, which has not been diagonalized so far. Therefore the mass eigenvalues are only known as elements of a single set without assignments to individual glueballs. The spectrum of the new $1^{++}$ glueball allows to separate one of the seven eigenvalue subsets and find one of the eigenvectors of the system.

Under the assumption that the glueball mass squared fits well a quadratic dependence we try to extract the remaining six $0^{++}$ states from the combined spectrum of~\cite{BHM2}. The assumption is certainly true for other glueballs studied previously. Nevertheless we can confidently extract only two more sub-spectra, which correspond to heavier states. This signifies that either the quadraticity of the spectra is violated by lighter glueballs, or numerics does not work so well for small eigenvalues in the coupled system.

This paper is organized as follows. In section~2 we study the general $\Ic$-even singlet vector ansatz. We show that apart from the $1^{++}$ dual to the $\uor$ current there is only one more vector glueball. In section~3 we compute its spectrum and find that it is degenerate with one of the seven scalars of~\cite{BHM2}. The Supersymmetric Quantum Mechanics (SQM) is used to derive the equation for the $0^{++}$ superpartner. In section~4 we analyze the full spectrum of $0^{++}$ states from~\cite{BHM2}. We conclude in section~5 with a remark on dual operators and proposal of the complete spectrum of $SU(2)\times SU(2)$ singlet $\Nc=1$ supermultiplets in the KS theory.


\section{New $\Ic$-even vector}
In this section we are going to consider the general ansatz for the $\Ic$-even vector fluctuations invariant under the global $SU(2)\times SU(2)$ symmetry of the KS solution.
To construct the ansatz we use the basis of $SU(2)\times SU(2)$ invariant one- and two-forms.\footnote{See~\cite{DMS} as well as \cite{KS,Tsimpis} for the background and coordinate conventions and the $\Ic$-values of the basis forms.} We represent all  4-dimensional vector fluctuations by bold face 1-forms, e.g. $\I\equiv \I_\mu(x,\tau)\d x^\mu$; assuming the notation $\d\I \equiv \d_4\I$ and transversality $\d *_4 \I=0.$ In what follows $x^\mu$ are the 4-dimensional coordinates, $\tau$ is a radial coordinate on the deformed conifold, $\epsilon$ is the deformation parameter and $g^5$ is a $\Ic$-even $SU(2)\times SU(2)$-invariant 1-form on the base of the conifold. Explicit form of various background functions used in the derivation can be found in the appendix.

First of all, we notice that there are no $\Ic$-even vector fluctuations of the 3-forms $F_3$ and $H_3$. This follows from the fact that they should flip the sign under the $\Ic$ conjugation while the only $\Ic$-odd basis forms are certain two-forms supported on the conifold.  The only singlet vector fluctuation of the metric has the form
\be
\label{genmet ansatz}
\delta (\d s^2) = \frac{2}{G^{55}}\,\left(\Vt\cdot g^5 + \Z\cdot \d\t\right)\,.
\ee
Taking into account this modification of the metric the general self-dual fluctuation of the 5-form reads
\begin{multline}
\label{gen5 ansatz}
\delta F_5 =  - \Wt \wg \d g^5\wg \d g^5 + \U\wg\d\t\wg\d g^5\wg g^5 + (\d\S+ *_4\d\T)\wg \d g^5\wg g^5  +
 (*_4\d\S -\d\T) \wg\d\t\wg \d g^5 -
\\ - h^{1/2}\sg G^{11}G^{33}(G^{55})^2 *_4 \U \wg \d g^5 + 2h^{1/2}\sg  (G^{11})^2(G^{33})^2\,*_4\Wt\wedge \d\t\wg g^5 -
\\ - \frac{\ell}{2}\,h^{1/2}\sg (G^{11})^2(G^{33})^2 \, *_4\Vt\wg\d\t\wg g^5\,.
\end{multline}
The ansatz (\ref{genmet ansatz}-\ref{gen5 ansatz}) is the most general ansatz for singlet $\Ic$-even vector fluctuations. The equations that are modified by the ansatz at the linear level are the following:

\begin{itemize}

\item The Bianchi identity $\d F_5 = H_3\wg F_3$ gives the equations
\be
\Wt' + \U  =  0\,,
\ee
\be
-\d\Wt + \d \S + *_4\d\T = 0\,,
\ee
\be
\d\U + \d\S' + *_4\d \T' =0\,,
\ee
\be
\d *_4 \d \T =0\,,
\ee
\be
\left(h^{1/2}\sg G^{11}G^{33}(G^{55})^2*_4\U\right)' + 2h^{1/2}\sg (G^{11})^2(G^{33})^2\left(*_4\Wt-\frac{\ell}{4}\,*_4\Vt\right) + \d *_4\d\S =0\,.
\ee

\item Three equations come from the components of the Einstein equation:\footnote{Here it is convenient to parameterize the background by the functions $p$, $x$ and $A$ introduced in~\cite{PT}. Expressions for them in the KS case can be found in the appendix.}
\be
-\delta R_{\mu\nu} = - \frac12\,\left(\pr_\mu\Z_\nu' + \pr_\nu\Z_\mu'\right) - (A'-6p')(\pr_\mu\Z_\nu + \pr_\nu\Z_\mu)=0\,,
\ee
\begin{multline}
-\delta R_{\mu\t} = \frac12\,\left(2A'' +4A'(2A'+x')+ m^2e^{-6p-x-2A}\right)\Z_\mu = \frac{\ell^2}{16}\,(G^{11})^2(G^{33})^2\Z_\mu +
\\ + \frac18\,\left(2{F'}^2G^{11}G^{33}+F^2(G^{11})^2+(1-F)^2(G^{33})^2\right)\Z_\mu\,,
\end{multline}
\begin{multline}
-\delta R_{\mu 5} = \frac12\,\Vt_\mu'' +\frac12\,\left(2A'-6p'+x'\right)\Vt_\mu' + \frac12\, m^2e^{-6p-x-2A}\Vt_\mu +
\\ + \frac12\,\left(-6p''-x''-2\left(2A'+x'\right)\left(6p'+x'\right)-2e^{-12p-4x}\right)\Vt_\mu = - \frac{\ell}{2}\, (G^{11})^2(G^{33})^2 \left(\Wt_\mu-\frac{\ell}{8}\,\Vt_\mu\right)+
\\ + \frac14\,\left(2{F'}^2G^{11}G^{33}+F^2(G^{11})^2+(1-F)^2(G^{33})^2\right)\Vt_\mu\,.
\end{multline}

\end{itemize}
The other supergravity equations remain unaltered at the linearized level.

Four of the five Bianchi identity equations can be solved to express everything in terms of the functions $\Vt$ and $\Wt$. The only one remaining equation is
\be
\label{general F5 eqn}
\Wt_\mu'' + \left(2\,\frac{K'}{K}-\frac{I'}{I}\right)\Wt_\mu' - \frac89\,\frac{\Wt_\mu}{K^6\sinh^2\t} + \mt^2\,\frac{I}{K^2}\Wt_\mu - \frac{2}{3}\,\frac{I'}{I}\,\frac{\V_\mu}{K^3\sinh\t}  =0\,,
\ee
where we have redefined
\be
\Vt = \frac{2^{1/3}3K}{I\sinh\t}\,\V\,,
\ee
and substituted for the 4-Laplacian, $*_4\d*_4\d \I = -\lpl \I $, its eigenvalue
\be
 \lpl \equiv m_4^2 = \frac{3\, \eps^{4/3}}{2\cdot 2^{2/3}}\, \mt^2 \,.
\ee

There is no non-trivial equation for $\Z$ from the Einstein equation. The only non-trivial equation that we obtain is the one involving $\V$ and $\Wt$,
\be
\label{gen metric eqn}
\V_\mu'' + \left(2\,\frac{K'}{K}-\frac{I'}{I}\right)\V_\mu' -\frac89\,\frac{\V_\mu}{K^6\sinh^2\t}  +  \mt^2\,\frac{I}{K^2}\,\V_\mu +\frac23\,\frac{I'}{I}\,\frac{\V_\mu-2\Wt_\mu}{K^3\sinh\t}  =0\,.
\ee
The general ansatz~(\ref{genmet ansatz}-\ref{gen5 ansatz}) leads to only two equations~(\ref{general F5 eqn}) and~(\ref{gen metric eqn}). In fact it reduces to the simplest generalization of the ansatz in~\cite{DM}. It is straightforward to diagonalize the system of equations~(\ref{general F5 eqn}) and~(\ref{gen metric eqn}).

Apparently there is a solution $\Wt_\mu=\V_\mu$. This is the solution found in~\cite{DM}, which corresponds to the $1^{++}$ glueball dual to the dimension $\Delta=3$ operator of the $\uor$ current in the gauge theory.
\be
\label{vector V}
\V_\mu'' +\left(2\,\frac{K'}{K}-\frac{I'}{I}\right)\V_\mu' -\frac{8}{9}\,\frac{\V_\mu}{K^6\sinh^2\t} + \mt^2\,\frac{I}{K^2}\,\V_\mu - \frac{2}{3}\,\frac{I'}{I}\,\frac{\V_\mu}{K^3\sinh\t} =0\,.
\ee
The other mode can be found by subtracting the first equation from the second one. Introducing a new function $\W_\mu=\V_\mu-\Wt_\mu$ we obtain
\be
\label{vector W}
\W_\mu'' +\left(2\,\frac{K'}{K}-\frac{I'}{I}\right)\W'_\mu -\frac{8}{9}\,\frac{\W_\mu}{K^6\sinh^2\t} + \mt^2\,\frac{I}{K^2}\,\W_\mu + 2\,\frac{I'}{I}\,\frac{\left(K\sinh\t\right)'}{K\sinh\t}\,\W_\mu  =0\,.
\ee
This vector mode corresponds to an operator of dimension $\Delta=7$. In the limit of the singular Klebanov-Tseytlin (KT) solution~\cite{KT} the equation~(\ref{vector W}) coincides with one of the vector equations found in~\cite{Krasnitz}.

\begin{table}[htb]
\begin{center}
\caption{\small First few $m^2$ eigenvalues and quadratic fit for the vector~(\ref{vector W})}
\label{tab W eigenvals}
\begin{tabular}{cc}
{\small a) in normalization of~\cite{BDKS}} &  {\small b) in normalization of~\cite{BHM2}}
\\
\begin{tabular}[b]{||c|c|c|c|c|c||}
\hline
4.38 & 7.08 & 10.3 & 14.2 & 18.5 & 23.5\\
\hline
29.0 & 35.1 & 41.8 & 49.1 & 56.9 & 65.3 \\
\hline
74.3 & 83.9 & 94.0 &  &  & \\
\hline
\end{tabular} &
\begin{tabular}[b]{||c|c|c|c|c|c||}
\hline
4.12 & 6.66 & 9.72 & 13.3 & 17.4 & 22.1\\
\hline
27.3 & 33.0 & 39.3 & 46.1 & 53.5 & 61.4\\
\hline
69.8 & 78.8 & 88.3 & & & \\
\hline
\end{tabular}
\\
$\mt^2 = 2.32 + 1.80\, n + 0.287\, n^2$ &  $m_{\text{BHM}}^2 = 2.17 + 1.70\, n + 0.269\, n^2$
\end{tabular}
\end{center}
\end{table}


\section{Superpartner}
\label{SQM}

We compute the spectrum of vector~(\ref{vector W}) numerically and present the results in table~\ref{tab W eigenvals}. The first 9 eigenvalues match with a good accuracy a subset of the eigenvalues computed by Berg, Haack and M\"uck (BHM) in~\cite{BHM2}. In table~5 of~\cite{BHM2} these eigenvalues are at positions $\#= 11, 19, 27, 34, 41, 48, 55, 62, 69$. For large values of the radial quantum number $n$ this subset has periodicity seven as expected. As we discuss below the degeneracy can be expected from the supermultiplet analysis. Thus the pseudovector $\W$ is the superpartner of one of the seven scalars in~\cite{BHM2}.

The scalar superpartner should correspond to either dimension $\Delta=6$ or $\Delta=8$. Starting with rewriting the equation~(\ref{vector W}) in the form
\be
\label{Shroedinger}
\psi'' - V \psi + \mt^2\,\frac{I}{K^2}\,\psi =0\,,
\ee
where
\be
V = -\frac12\, \frac{(I/K^2)''}{I/K^2} + \frac{3}{4}\,\frac{{(I/K^2)'}^2}{(I/K^2)^2}+\frac89\,\frac{1}{K^6\sinh^2\t} - 2\,\frac{I'}{I}\,\frac{(K\sinh\t)'}{K\sinh\t}\,,
\ee
and applying the SQM transformation following~\cite{DM} and~\cite{BDKS} we obtain the superpartner equation:
\be
\label{superpartner eqn}
\psi''_s - V_s \psi_s + \mt^2\,\frac{I}{K^2}\,\psi_s =0\,,
\ee
where the potential $V_s$ is
\be
\label{superpartner potential}
V_s= -\frac{8}{3}\,\frac{\sinh{2\tau}}{\sinh{2\tau}-2\tau} + \frac{160}{9}\,\frac{\sinh^4{\tau}}{(\sinh{2\tau}-2\tau)^2}\,.
\ee
The asymptotic form of the superpartner equation at the boundary $\t\to\infty$ is
\be
\psi''_s - \frac{16}{9}\, \psi_s =0\, ,
\ee
which allows to establish the dimension of the dual operator to be $\Delta=2+\sqrt{4+(16-4)}=6$.

It is also interesting to notice that with the help of this trick with SQM we find one of the seven eigenvectors of the complicated system in~\cite{BHM2}, which means that its rank can be reduced by one. We argue below that it is reasonable to expect that the rank can be further reduced at least by two, i.e. two more eigenvectors can be decoupled.


\section{Structure of the lightest scalar spectrum}
\label{Berg spectrum}
In the works of BHM~\cite{BHM1,BHM2} an impressive effort was made to derive the spectrum of all $0^{++}$ glueballs in the KS background. As the result a system of linearized equations was computed and the spectrum of glueballs was found numerically. However the linearized system appeared to be coupled and complicated enough to discourage further efforts to diagonalize it. As a consequence in the spectrum computed in~\cite{BHM2} the eigenvalues are mixed (at least the lightest ones) without the assignment to the individual glueball towers. As we have shown above, there should be a way to decouple at least one equation from the other six. We have presented the corresponding eigenvector~(\ref{superpartner eqn}).

Comparing the results on the spectrum of $1^{++}$ and the table of $0^{++}$ eigenvalues we have confirmed that the vector belongs to a massive supermultiplet together with one of the scalars. We have therefore isolated a subset of eigenvalues from the full spectrum of $0^{++}$. Moreover we have noticed that the subset fits very well a quadratic form. This is also true about other glueball spectra, e.g. the ones of the $\Ic$-odd states. Therefore one could try to disentangle the eigenvalue system of BHM assuming that the remaining glueballs have a quadratic spectrum with a sufficient accuracy.\footnote{In fact such an attempt was done in~\cite{BHM2}, but apparently none of the seven towers identified there describe the $1^{++}$ multiplet. In what follows we will try to somewhat improve those results.}

As noticed in~\cite{BHM2} for large values of $m_{\rm BHM}^2$ the spectrum shows some periodic pattern with expected periodicity seven. We have employed the following approach. First we have obtained an initial fit for few (3-4) points, which appeared periodically, and afterwards added the points predicted by the trial fit and checked the stability of the trial fit with respect to such an extension of the data set.

Despite the fact that this approach predicts several more points with a nice accuracy, in general it fails for light eigenvalues. Besides the known subset of the $1^{++}$ multiplet we have only found two more stable fits. Those correspond to the following subsets in table 5 of~\cite{BHM2}:
\begin{itemize}
\item $\# = 7,12,18,25,33,40,47,54,61,68,75$ can be fitted with the parabola
\be
\label{xtra glueball1}
m_{\rm BHM}^2 = 0.269 n^2 + 1.054n + 1.01\,,
\ee
\item $\# = 10,16,23,30,37,44,51,58,65,72$ can be fitted by
\be
\label{xtra glueball2}
m_{\rm BHM}^2 = 0.271 n^2 + 0.787n + 0.619\,.
\ee
\end{itemize}

\begin{figure}[htb]
\begin{center}

\epsfxsize=3.9in \epsffile{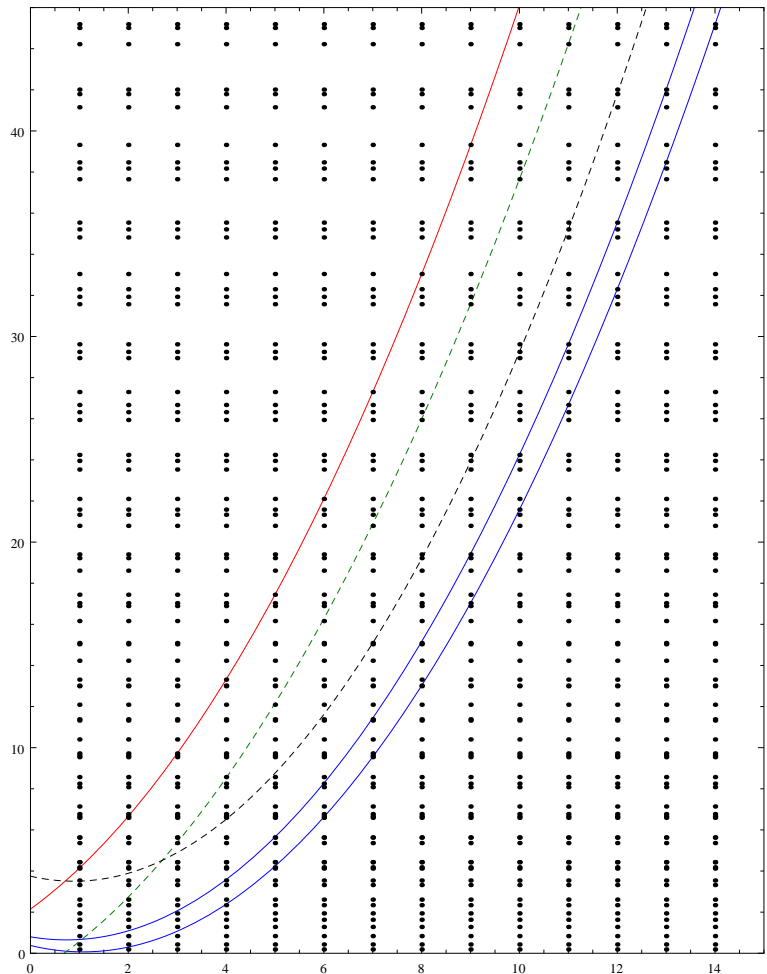}

\end{center}
\vspace{-0.6cm}
\caption{\small A fitting of the $0^{++}$ spectrum found in~\cite{BHM2}. Above each value of $n$ on the horizontal axis we put the whole spectrum of $m_{\rm BHM}^2$ (a vertical line of dots). The plot shows the known set from table~\ref{tab W eigenvals} (red), two stable fits (blue), accidental good fit (green, dashed) and an example of not so good fit. The lightest eigenvalues in a particular fit do not necessarily correspond to $n=1$. }
\label{fig fitting}
\end{figure}

In figure~\ref{fig fitting} we show how the trial fits work in several examples. In case of stable fits it is enough to fit by 3 points. We also show several unstable fits. One of them is accidentally good if one considers just a 3-point fit. However it gets distorted after adding more points. Moreover, the $n^2$ coefficient for this fit unusually deviates from the value $0.27$, which should be more or less universal for all glueballs~\cite{DM}.

The reason that two glueball towers~(\ref{xtra glueball1}) and~(\ref{xtra glueball2}) are special is perhaps due to the fact that they are heavier and correspond to operators of larger dimensions ($\Delta=7,8$). It has been demonstrated in~\cite{BHM1} that in the KT limit these glueballs decouple from the remaining system. The lighter four glueballs correspond to the dimension $\Delta=3,4$ operators such as the gluino bilinear and trace of energy-momentum tensor. These operators have large mixing and it is natural that corresponding glueballs also mix. Based on the result of the above analysis and KT limit result one might expect that equations for glueballs (\ref{xtra glueball1}) and (\ref{xtra glueball2}) can be decoupled from the full system.


\section{Discussion}
In this work we have considered the most general ansatz for $\Ic$-even $SU(2)\times SU(2)$ singlet vector fluctuations of the type IIB supergravity over the \KS solution. The ansatz reduces to two non-trivial independent fluctuations, one of which is the pseudovector dual to the $\uor$ current operator. The second pseudovector is dual to an operator with the dimension $\Delta=7$ and zero ${\cal{R}}$-charge.

To determine the operator dual to the second $1^{++}$ one can scan the 5-dimensional superconformal multiplet classification~\cite{CDDF} of the conformal theory on the conifold~\cite{KW}. A $SU(2)\times SU(2)$ singlet vector with dimension 7 only enters the type II Vector Multiplet. This long multiplet is the analog of the reducible real superfield $V$; and the pseudovector is its $\theta\sigma^\mu\bar{\theta}$ component. In superfield notations the operator corresponding to the type II Vector Multiplet is the operator $\tr W^2\bar{W}^2$~\cite{Apreda}. The $\theta\sigma^\mu\bar{\theta}$ component of this operator is
\be
O_{1^{++}}= \tr \big\{\lambda,F_{\alpha\beta}\big\}\sigma^{\alpha}\big\{\bar{\lambda},{F^{\beta}}_\mu\big\} + \tr \big\{\lambda,\tilde{F}_{\alpha\beta}\big\} \sigma^{\alpha}\big\{\bar{\lambda},{{{\tilde{F}}^{\beta}}}_{~~\mu}\big\}\,.
\ee

We have computed the spectrum of the new $1^{++}$ glueball and found it to be degenerate (with a very good accuracy) with a subset of eigenvalues of the combined spectrum of the seven $0^{++}$ glueballs found in~\cite{BHM2}. This signifies that the $1^{++}$ is in the same 4-dimensional $\Nc=1$ multiplet with one of the seven $0^{++}$. One could also anticipate this after analyzing the content of the superconformal Vector Multiplet~II. In section~\ref{SQM} we have established the dimension of the superpartner $0^{++}$ to be $\Delta=6$. Therefore it corresponds to the lowest component of the $\tr W^2\bar{W}^2$ operator, which is a 4-fermion operator
\be
O_{0^{++}}= \tr \lambda\lambda\bar{\lambda}\bar\lambda\,.
\ee

\begin{figure}[htb]
\begin{center}

\epsfxsize=4.5in \epsffile{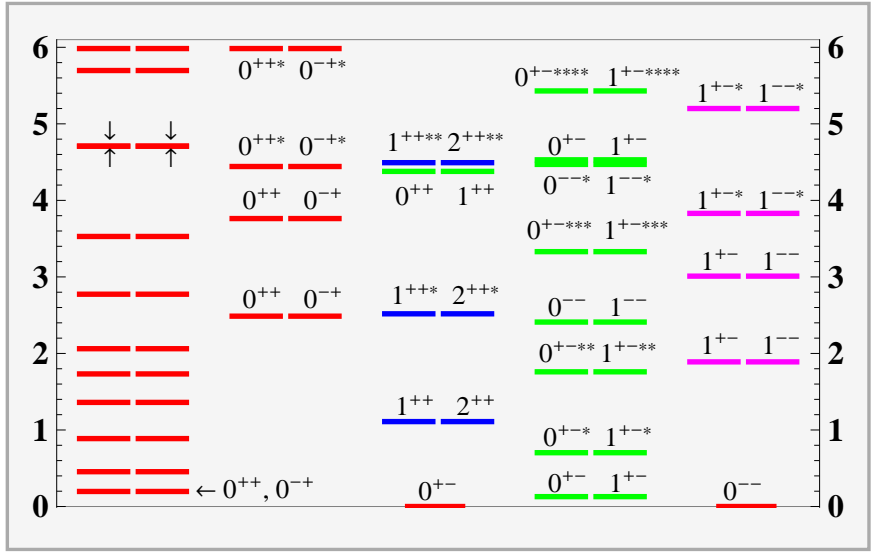}

\end{center}
\vspace{-0.6cm}
\caption{\small The complete spectrum of the lightest ($\mt^2<6$) $SU(2)\times SU(2)$ singlet glueball multiplets of the \KS theory. The multiplets are represented by their bosonic members. Different colors correspond to: the graviton (blue), gravitino (magenta), vector (green) and scalar (red) multiplets. The spectra of the six scalar multiplets in first two columns were computed in~\cite{BHM2} and their full structure is not yet clear (no labels for the states in the first column). Double arrows indicate two degenerate states.}
\label{fig f1}
\end{figure}

The fact that there are no other vector states means that the only missing states from the spectrum are the pseudoscalars $0^{-+}$, which should necessarily form scalar multiplets with the remaining 6 scalars in~\cite{BHM1,BHM2}. Given that the $\Ic$-odd sector was fully studied in~\cite{BDKS,DMS} and the spectrum of $0^{++}$ is known from~\cite{BHM2}, we should know the full spectrum of $SU(2)\times SU(2)$ singlet glueballs in the KS theory. Here in figure~{\ref{fig f1}} we propose the structure of the full spectrum.

In figure~\ref{fig f1} each supermultiplet is represented by its bosonic members. The multiplets are labeled by the highest spin members except for scalar multiplets. There are two massless multiplets -- scalar multiplets containing only one bosonic spin 0 state. The massive sector contains one tower of graviton multiplets, two towers of gravitino multiplets, four towers of vector multiplets and six towers of scalar multiplets.

Some of the scalar multiplets in figure~\ref{fig f1} appear without labels. As we discussed above one still has to disentangle six individual towers from the combined spectrum of $0^{++}$ in~\cite{BHM2}. Based on our analysis in section~\ref{Berg spectrum} we have extracted two heavier subspectra out of six, but the procedure did not work as well for the remaining four that appear in the first column.

Since the spectrum of the four lightest $0^{++}$ deviates from the quadratic form it is also possible that the lightest eigenvalues are affected by a numerical error. As an independent check of the lightest eigenvalues one could try to compute the spectrum of the pseudoscalar superpartners, which would be a topic of a separate work~\cite{finale}.

\vspace{0.5cm}
We are indebted to S.~Cremonesi, A.~Dymarsky, I.~Klebanov, A.~Solovyov and A.~Vainshtein for useful discussions. This work was partially supported by the Israeli Science Foundation center of excellence, the Deutsch-Israelische
Projektkooperation (DIP), the US-Israel Binational Science Foundation (BSF), the German-Israeli Foundation (GIF), the RFBR grant 07-02-01161 and the grant of the President of the Russian Federation for the Support of Scientific Schools NSh-3035.2008.2.

\appendix
\setcounter{equation}{0}
\renewcommand\theequation{A.\arabic{equation}}
\section*{Appendix}
Here we present explicit expressions for the background functions used in our derivation.
\begin{align}
h &= 4 \cdot 2^{2/3} \eps^{-8/3} I\,,\\
I &=
\int_\tau^\infty \d x\, \frac{x\coth x-1}{\sinh^2 x}\, (\sinh (2x) - 2x)^{1/3} \,,\\
K &= \frac{(\sinh 2\t-2\t)^{1/3}}{2^{1/3}\sinh\t}\,,\\
G^{11}  &= \frac{2}{\epsilon^{4/3} K\sinh^2 (\tau/2) h^{1/2}} \,,\\
G^{33}  &= \frac{2}{\epsilon^{4/3} K\cosh^2 (\tau/2) h^{1/2}} \,,\\
G^{55}  &= \frac{6\, K^2}{\epsilon^{4/3} h^{1/2}} \,,\\
\sg &= \frac{\eps^4}{96}\, \h{1/2} \sinh^2\t \,,\\
\ell &= \frac{\tau\coth\tau - 1}{2\sinh^2\t}\left(\sinh 2\tau -2\tau \right)\,,\\
A &= - \frac14\,\ln h\,,\\
e^{6p+2x} &=\frac32\,(\coth\tau-\tau{\rm csch}^2\tau)\,,\\
e^{2x} &=\frac{1}{16}\,(\sinh\tau\cosh\tau - \tau)^{2/3}h\,.
\end{align}


\end{document}